# LaB$_6$ aided spontaneous conversion of bulk graphite into carbon nanotubes at normal atmospheric conditions

Shalaka A. Kamble [a], Soumen Karmakar [b, *], Somnath R. Bhopale [a, c], Sanket D. Jangale [a], Neha P. Gadke [a], Srikumar Ghorui [d], S. V. Bhoraskar [a], M. A. More [a], V. L. Mathe [a, *]

[a] Department of Physics, Savitribai Phule Pune University, Ganeshkhind Pune 411007, Maharashtra, India
[b] Department of Physics, Birla Institute of Technology, Mesra, Off-Campus Deoghar, Deoghar 814142, Jharkhand, India
[c] Present address: School of Physics, IISER Thiruvananthapuram, Vithura, Kerala 695551, India
[d] Homi Bhabha National Institute and Thermal Plasma Technologies Section, Bhabha Atomic Research Centre, Trombay, Mumbai-400085, India

* Corresponding author:
 E-mail address: skarmakar@bitmesra.ac.in (S. Karmakar), vlmathe@physics.unipune.ac.in (V. L. Mathe)

**ABSTRACT**

Herein, we report a case study in which we saw the spontaneous conversion of commercial bulk graphite into LaB$_6$ decorated carbon nanotubes (CNTs) under normal atmospheric conditions. The feedstock graphite was used as a hollow cylindrical anode filled with LaB$_6$ powder and partially eroded in a DC electric-arc plasma reactor in pure nitrogen atmosphere. An unusual and spontaneous deformation of the plasma-treated residual anode into a fluffy powder was seen to continue for months when left to ambient atmospheric conditions. The existence of LaB$_6$ decorated multi-walled CNTs at large quantity was confirmed in the as-generated powder by using electron microscopy, Raman spectroscopy and x-ray diffraction. The as-synthesized CNT-based large-area field emitter showed promising field-emitting properties with a low turn-on electric field of ~1.5 V $\mu$m$^{-1}$, and a current density of ~1.17 mA cm$^{-2}$ at an applied electric field of 3.24 V $\mu$m$^{-1}$.

1. **Introduction**

Since the discovery of carbon nanotubes (CNTs) [1], these novel allotropic forms of carbon have gained significant scientific attention owing to their exceptional physical, electrical, optical, and mechanical properties [2,3]. They have high tensile strength, high electrical and thermal conductivities, chemical stability, and excellent electrical properties. Owing to these properties, CNTs have become potent candidates for widespread use in the fields of electrical, electronic, optical, mechanical, biomedical applications, sensors and many more [4–6].

Owing to the continuously increasing applications of CNTs, researchers are motivated to explore various avenues for synthesizing CNTs with the aim of achieving absolute control over their yield, cost, purity, structure, alignment, and properties. To achieve these goals, techniques such as arc discharge, chemical vapor deposition, laser ablation, plasma-assisted synthesis, liquid-phase growth, pyrolysis, fluidized bed reactions, and solid-state growth remain popular and widely explored [7].

However, most of the outstanding properties of CNTs reported so far are generally exhibited by CNTs, which are typically produced by an arc discharge method. This is because an electric arc is known to aid in the growth of pristine CNTs owing to its elevated temperature, which is comparable to the sublimation point of graphite. In an arc discharge method, a DC electric arc is generally ignited in an inert gaseous environment between two coaxial cylindrical graphite rods separated by a distance typically of the order of millimeter. During arcing, the anode is sublimated in an inert gas atmosphere, and a columnar deposit is formed *in situ* on the anode-facing surface of the cathode. Multi-walled CNTs (MWCNTs) can be found inside these cathode deposits if the conditions are favorable [8]. On the other hand, if a graphite crucible is filled with suitable catalyst(s) and then vaporized by an arc, it is possible to obtain single-walled CNTs (SWCNTs) in the soot, which mainly gets deposited on the walls of the arc plasma reactor [9]. The generation of CNTs stops as soon as the arcing is switched off. After arcing, if the electrodes are left to the ambient environment, no further structural deformation generally occurs.

However, for the first time, we have seen catalyst-driven *ex situ* spontaneous growth of crystalline CNTs from an arc-treated graphite anode, when left to normal atmospheric conditions, if certain conditions are met. Herein, we report our observations that not only may invoke an interest in investigating the growth mechanism of CNTs from bulk graphite but may also open new avenues for the commercial synthesis of CNTs in a very economical, energy-efficient, reproducible, and pollution-free manner, completely avoiding any sophisticated instrumentation. Among the many possible applications of as-synthesized CNTs, we explored the possibility of using these CNTs as field emitters owing to their high aspect ratio and conical caps.

2. **Experimental details**

A double-walled, water-cooled stainless steel reactor chamber was used for the arcing process. A schematic of the reactor and electrode configuration is shown in Fig. 1.

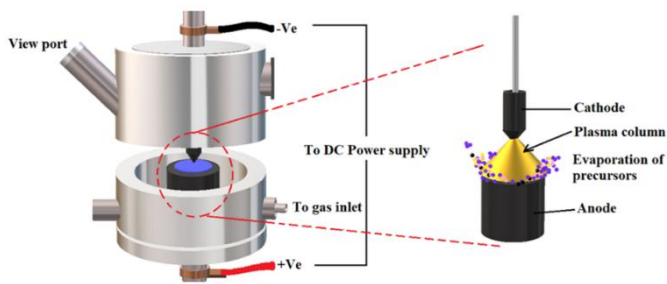

**Fig. 1- Schematic of the arc plasma reactor and its operation.**

Two coaxial cylindrical electrodes consisting of 99.99% pure spectroscopic pyrolytic graphite rods were used as electrodes. The anode was partially drilled along its axis to form a crucible in which 99.95 % pure microcrystalline $LaB_6$ powder (manufactured by Sigma-Aldrich Chemicals USA) was filled before arcing as mentioned elsewhere [10]. The operational parameters of the reactor, which were kept during the synthesis runs, are listed in Table 1.

**Table 1 - Operational parameters of the arc reactor chamber.**

| | |
|---|---|
| Diameter of the anode | 10 mm |
| Diameter of the cathode | 30 mm |
| Diameter of the anode-hole | 28 mm |
| Operating chamber pressure | 500 ± 10 Torr |
| Arc-current | 150 ± 5 A |
| Arc-voltage | 20 ± 2 V |
| Typical run-time of each operation | ~5 min |

The arcing process was performed in an inert atmosphere. The reactor was first evacuated to a base pressure of $10^{-3}$ Torr with the help of an oil rotary vacuum pump. The chamber was purged with buffer gas and re-evacuated several times to free the inner chamber atmosphere of oxygen. Finally, the chamber was filled with buffer gas to the level of the operating chamber pressure and arcing was carried out using a traditional 'touch and pull' method. 99.99% spectroscopic pure $N_2$ was used as the buffer gas. The operational parameters were selected such that the $LaB_6$ powder could melt and some of its residue remained intact inside the anode crucible.

$LaB_6$ is chemically inert at room temperature. However, it oxidizes at elevated temperatures and the resulting $La_2O_3$ is highly reactive and hygroscopic in nature. To avoid such chemical reactions the reactor chamber was first allowed to cool to the ambient atmospheric temperature. The chamber pressure was then brought back to the atmospheric level before opening the chamber to dismount the anode.

The dismantled anode, with the $LaB_6$ residue when left to the atmospheric conditions was found to undergo spontaneous structural deformation. With time, the arc-treated surface of the anode was covered with a soft Colliflower-like structure that could easily be removed and characterized. After collection, the Colliflower-like sample was first mechanically homogenized by a granite-made pestle and mortar and then characterized with the help of Raman spectroscopy, electron microscopy, x-ray diffraction (XRD) and field emission microscopy (FEM).

We recorded the Raman spectra at room temperature by a Renishaw inVia™ spectrometer, where a He-Ne laser beam with 1 mW power and 1 μm diameter was used. Raman spectra were recorded in the range of Raman shift from 100 cm$^{-1}$ to 3000 cm$^{-1}$.

Electron microscopic investigations were carried out using field-enhanced scanning electron microscopy (FESEM), transmission electron microscopy (TEM) and high-resolution TEM (HRTEM). FESEM, TEM, and HRTEM investigations were carried out using an FEI Nova Nano SEM 450 microscope, a Philips CM 200 microscope, and a Tecnai G$^2$, F30 microscope, respectively. To carry out the electron microscopic investigation, ~5 mg of each powdered sample was ultrasonicated in ~ 30 mL toluene for ~30 min in glass test tubes and then drop-casted on carbon-coated lacey copper grids before drying under an IR lamp.

The crystallographic structures of the as-synthesized samples were analyzed by x-ray diffraction (XRD) using an x-ray diffractometer (Bruker, Model D8), where Cu-K$_α$ radiation was used as x-ray.

Field emission (FE) properties of $S_6$ were investigated using an indigenously developed in-house experimental facility. To perform the FE measurements, the samples were mounted on carbon tapes and used as the cathode, while phosphor-coated ITO glass was used as the anode. The operational parameters of the FE setup were the same as those described elsewhere [11]. The FE current was recorded as a function of the applied voltage using a multimeter (Rishabh, 14 S series) and a DC high-voltage supply (Spellman, USA). A digital camera (Canon SX150IS) was used to record the FE micrographs.

### 3. Results and discussion

Fig. 2 shows typical photographs of the dismantled anode within six months. Initially, we saw a shiny grey coating on the plasma-treated surface of the anode (Fig. 2a). There was no visible change in the morphology of the anode surface for the first couple of days. However, the surface became dusty no sooner than 48 h of arcing and Colliflower-like eruptions then started evolving naturally on this surface when the anode was exposed to ambient atmospheric conditions over a long time (Fig. 2c). We collected the sample in batches from the anode surface once a Colliflower-like deformation developed. With the removal of this deformed structure, another fresh Colliflower-like structure grew on the anode surface (like Fig. 2a) if it had some remnants of $LaB_6$.

However, before we discuss our findings related to a particular batch, we would like to mention that we explored various conditions for the spontaneous conversion of the eroded anode. However, we found such transformation to be either negligible or nil subject to the following conditions: (a) when Ar was used as the plasma forming gas, (b) if the eroded anode was kept under high vacuum conditions (pressure ~$10^{-6}$ Torr), or in an inert ambience at atmospheric pressure; (c) if the $LaB_6$ containing anode was kept inside a closed glass jar with normal atmospheric conditions inside, (d) if the plasma-melted $LaB_6$ was mixed with commercial graphite powder and left to the ambient atmosphere; (e) if the shiny layer of $LaB_6$ was removed completely from the anode surface immediately after taking the anode out of the reactor chamber; (f) if after the synthesis run $LaB_6$ layer was scraped from the anode surface and placed on a piece of bulk graphite, which was not arc-treated; and (g) if the powders of $LaB_6$ and commercial graphite were mixed together, plasma-treated and then left to the ambient atmosphere.

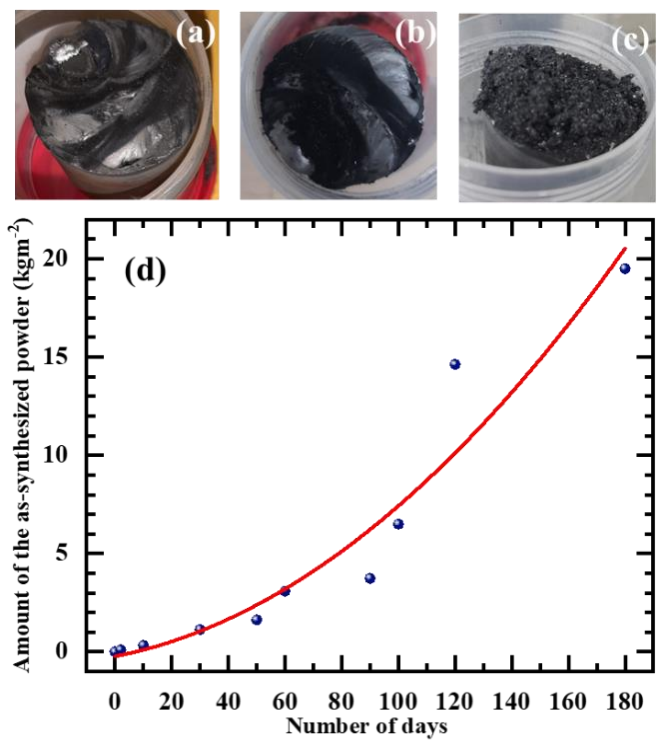

**Fig. 2 – Photographs showing the morphological changes of the anode surface (a) just after the arcing, (b) ten days after the arcing. (c) Photograph of the anode on which the deformation had taken place between four to six months after the arcing. (d) The rate of conversion of graphite anode into Colliflower-like structure. The mass measurements were performed using a Mettler Toledo digital balance. The solid red line is a guide to the eye.**

It is also noteworthy that we saw a similar conversion of graphite (as shown in Fig. 2a-c), when we replaced $LaB_6$ with other rare-earth hexaborides and mixed hexaboride powders (e.g., $CeB_6$, $(GdLa)B_6$, $(GdCe)B_6$) during the arcing process. However, in this communication, we report our observations related to $LaB_6$ only.

Fig. 2d shows the rate of deformation (measured in terms of the mass of the deformed anode-structure) of the anode, with an effective active area of $6.15 \times 10^{-4}$ $m^2$ over the time. The figure shows that the overall rate of such conversion did not remain constant with time, but increased rapidly after two months of arcing, following a nonlinear pattern. The scattering of data points from the trend-line (the red line in the figure) may be attributed to the fluctuations in the weather-parameters that are likely to affect the rate of anode deformation.

Although we obtained comparable results related to different batches of the sample we collected, here we report our observations for the batch that grew in between 2 – 6 months from arcing. For the rest of the manuscript, this sample is called $S_6$. Electron microscopy analysis of $S_6$ collected after six months of arcing revealed that $S_6$ had many thread-like structures and a few particles (Fig. 3a). While analyzing the morphology of such threads by TEM, we found that many of the threads underwent antenna-like growth (showed by arrows in Fig. 3b), thereby increasing their effective aspect ratio.

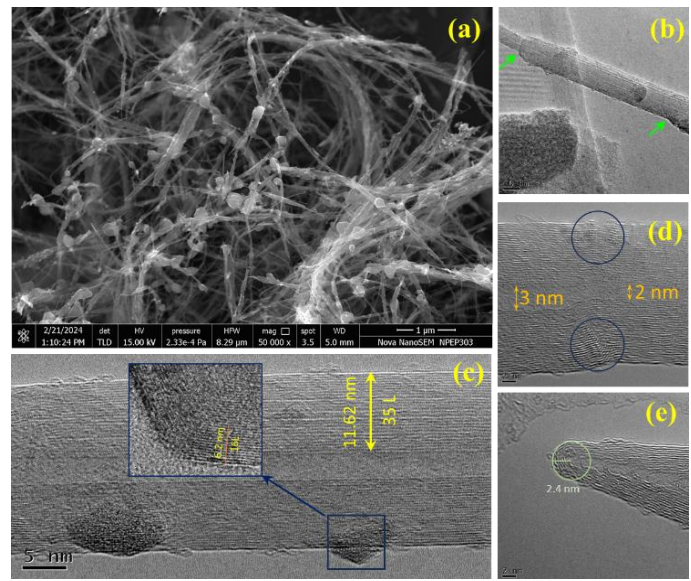

**Fig. 3 – Characteristics of $S_6$ as revealed by electron microscopy. (a) Typical FESEM micrograph of the sample, (b) typical TEM micrograph of a CNT in which the inner cylinders grew more than the outer cylinders along the length of the tube, and (c-e) typical HRTEM micrographs of the as-synthesized CNTs.**

The fine structures of these nanothreads were further revealed by HRTEM (Fig. 3c-e). Fig. 3c shows that the nanothreads are hollow concentric cylinders with some nanoprotrusions distributed randomly over the threads. The interwall separation of such cylinders was found to be ~ 3.4 Å, which is remarkably close to the interplanar separation of the basal planes of graphite. On the other hand, the nano protrusions onto these tubes were identified as $LaB_6$ nanoparticles, whose (100) crystallographic planes were clearly visible in the corresponding HRTEM micrographs (inset of Fig. 3d). Fig. 3d thus confirms that the synthesized nanothreads are nothing but $LaB_6$ decorated CNTs. Upon examining such CNTs by HRTEM, we found that many of them were poorly crystalline with wavy walls and variable inner diameters, as highlighted by circles in Fig. 3d. Some of the CNTs were found to have very sharp conical tips (Fig. 3e) like the ones often seen in the case of arc-generated CNTs [12].

Fig. 4a shows a typical XRD pattern of $S_6$, in which the diffraction maximum, due to the (002) basal planes of graphite, is visible along with strong signatures of nanocrystalline $LaB_6$ [8,14]. No other crystalline species of La was seen in any of the recorded XRD spectra. This clearly shows that there was no chemical reaction of $LaB_6$ with the nitrogen plasma during arcing and the plasma treated sample also did not undergo any further chemical change, whatsoever. The value of the interplanar separation of the basal planes of graphite ($d_{002}$), seen in this figure matches well to both the values estimated from the typical HRTEM micrograph of $S_6$ (Fig. 3d) and the arc-generated CNTs [8]. From Fig. 4a we also see that the intensity ratio of the (002) peak of graphite to that of the most intense (110) peak of $LaB_6$ is ~ 0.24 despite $S_6$ being a mainly carbonaceous sample. Fig. 4a thus shows that the degree of crystallinity of the graphitic content of $S_6$ is poorer than that of coexisting $LaB_6$. This conclusion is supported by Fig. 3d. Moreover, the absence of (101) peak of graphite and the noise present in the XRD spectrum confirm that the crystalline graphitic species present in $S_6$ are turbostratic and have considerable amounts of structural defects [15].

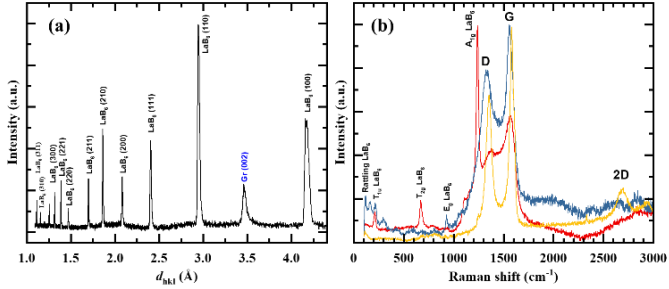

Fig. 4 – (a) Powder XRD pattern of $S_6$. (b) Typical normalized Raman spectra of $S_6$ collected from various locations of the sample. The $LaB_6$ related Raman peaks were identified with the help of the existing literature [13].

The conclusions made based on the electron microscopy and XRD analysis of $S_6$ were further supplemented by the Raman spectroscopy. As $S_6$ was inhomogeneous with respect to the diameter of the used laser spot, different Raman spectra were obtained at various locations on the sample. While recording the Raman spectra of $S_6$, we obtained mainly three types of spectra, as highlighted in Fig. 4b. In all the recorded spectra two prominent peaks near 1580 cm$^{-1}$ and 1340 cm$^{-1}$ were common. These are the G and D bands, respectively. G band arises due to in-plane vibration of hexagonal graphite lattice, while the D band is a signature of the defects and disorders present the graphitic structures [8]. D band may also arise due to an interaction of the laser with the edges of the small crystallites of graphite [16]. In the first type of Raman spectra (the yellow one in Fig. 4b) the G band appeared at 1581 cm$^{-1}$ and the intensity ratio of the G band ($I_G$) to the D band ($I_D$) was higher than the other types of Raman spectra. This type of spectrum can best be attributed to the pyrolytic graphite matrix that underwent structural transformation in presence of $LaB_6$. The red spectrum in this figure is dominated by the large microparticles of $LaB_6$, coated with graphitized carbon and shows the most profound $A_{1g}$ Raman active mode near 1235 cm$^{-1}$ with some other lower order peaks. This type of carbonaceous $LaB_6$ particles is very common in high temperature plasma systems in which graphite is used as an anode [17]. On the other hand, in the blue type of spectrum some prominent lower order peaks are visible in the range from 100 cm$^{-1}$ to 300 cm$^{-1}$ which are attributed to the phonon vibrations of $LaB_6$ nanoparticles, whose typical particle size is 5 nm or less [13]. In both the red and blue types of spectra in Fig. 4b, the $I_G/I_D$ ratio was seen to be much lower, and the G peak was seen to downshift by ~20 cm$^{-1}$ with respect to the yellow type. Such downshift was also seen in the case of SWCNTs with doping [18]. We guess that the observed downshift of G peak position in our case might also be due to the presence of $LaB_6$ nanoprotrusions (with less than 5 nm size) in the as-synthesized MWCNTs (as seen in Fig. 4c).

However, more studies are needed to verify this. The poor crystallinity of the as-synthesized MWCNTs, as is revealed in Fig. 3d, is possibly the reason behind the intense disorder-induced D band in the blue type of Raman spectra we saw in the case of $S_6$. The vanishing intensity of the 2D band in this type of Raman spectra is possibly due to an enhanced electron scattering with the defects [19] present in the as synthesized CNTs.

The typical dimension, morphology of the tips (Fig. 3e) and the defects present in the as-synthesized CNTs (Fig. 3d) encouraged us to study the large area FE properties of our sample, as both the defects and $LaB_6$ nanoparticles present in a graphite-like system (like the one seen in Fig. 3c) aid in lowering the work function ($\varphi$) of the material [20,21] that helps in attaining cold filed emission at a comparatively low applied electric field ($E$).

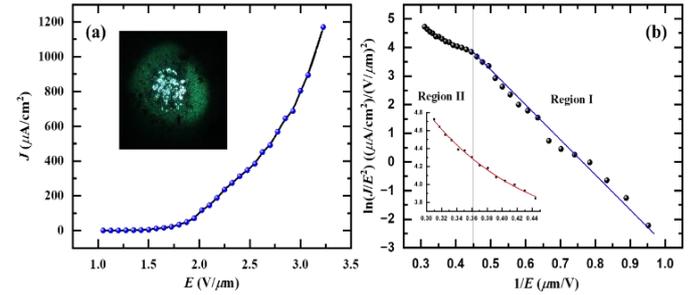

Fig. 5 – Field emission properties of $S_6$ at room temperature. (a) The variation of $I$ as a function of $V$. The inset shows the field emission micrograph. Multiple spots in this micrograph show the distribution of our multi-tip field emitters. (b) The corresponding F-N plot. The blue line is the linear fit to the data points corresponding to regions I. The red line in the inset shows the nonlinear nature of the F – N plot in region II.

Fig. 5 shows the outcome of our FE experiment. From the observed variation $J$ (the emission current density computed over the area of the cathode) as a function $E$ (Fig. 5a) we estimated the turn-on field to be 1.5 V $\mu$m$^{-1}$, which was needed to obtain $J = 1$ $\mu$A cm$^{-2}$. The threshold electric field (field needed to generate $J=100$ $\mu$A cm$^{-2}$) was found to be 1.9 V $\mu$m$^{-1}$. A comparison of our $J – E$ data with the one obtained for nano $LaB_6$ particles [17,22] shows that the CNTs present in $S_6$ are most likely the primary contributors in the observed FE. However, the properties of our field emitter could be understood better by analyzing the corresponding Fowler-Nordheim (F – N) plot (as shown in Fig. 5b), the simplified equation [22] of which may be written as:

$$\ln\left(\frac{J}{E^2}\right) = \left(\frac{-b\varphi^{3/2}}{\beta}\right) \times \frac{1}{E} + \ln\frac{a\beta^2}{\varphi} \quad (1)$$

Here, $a$ and $b$ are constants with values 1.54434×10$^{-6}$ A eV V$^{-2}$ and 6.83089×10$^9$ V eV$^{-3/2}$ m$^{-1}$ [21], $\varphi$ is expressed in eV, and $\beta$ is the field enhancement factor. Eq. (1) is an equation of straight line $y = mx + c$, where $x = \frac{1}{E}$, $y = \ln\left(\frac{J}{E^2}\right)$, $c = \ln\frac{a\beta^2}{\varphi}$, and

$$m = \left(\frac{-b\varphi^{3/2}}{\beta}\right) \quad (2)$$

The F – N plot, obtained from our $J – E$ data (Fig. 5a), did not show a constant linearity throughout, but followed a linear variation only in the range $0.95\ \mu mV^{-1} \geq E^{-1} \geq 3.24\ \mu mV^{-1}$, as highlighted in Fig. 5b. When $E^{-1} \leq 3.24\ \mu mV^{-1}$, the FN plot deviated from linearity and followed a nonlinear variation (inset of Fig. 5b). Most commonly, in the case of CNT-based field emitters $m$ is found to increase at higher $E$ [21], an effect which has been attributed to the degradation of CNT emitters [23]. However, our sample showed a sudden decrease in $m$ while going from region I to II in Fig. 5b, like Q-carbon emitters [24]. A decrease in $m$ while going from region I to region II has often been attributed to an increase in $\beta$ assuming $\varphi$ to be a constant throughout [24]. However, this might be an oversimplification of our case ignoring the involvement of Joule heating and heating of emitter tips due to ion bombardment by residual gases when $E$ is sufficiently large. Owing to the positive temperature coefficient of electrical conductance [25] Joule heating dominates in CNTs with the increase in $E$ [26] thereby lowering the value of $\varphi$. The observed change in the slope of the F – N plot while going from region I to region II (Fig. 5b) may therefore be a sign of a changeover from the cold FE (CFE) to the thermionic FE, as has been discussed in the literature [27]. For $S_6$, the critical electric field at which this changeover took place was estimated to be 2.24 V $\mu m^{-1}$. Though we could not determine the exact value of $\varphi$ for our field emitter directly, it is obvious that it must lie between the corresponding values of pristine CNT ($\varphi_{CNT} = 5.05\ eV$) [28] and pure LaB$_6$ ($\varphi_{LaB_6} = 2.07\ eV$) [29]. Using these values and Eq. (2) we obtain $1739 < \beta < 6626$, which is a promising figure for considering our LaB$_6$ decorated CNTs as possible field emitters.

It is to be noted that we have examined all the samples obtained from the different batches of the as-generated Colliflower-like structures and did not find any substantial differences in the characteristic features of $S_6$ as reported in this paper. We have also noticed that once removed from the anode surface, none of the sample underwent any further structural or morphological changes with time. It is quite clear from our analysis that with the removal of a batch, the quantity of LaB$_6$ available to the next batch is less; however, the yield of production increased (Fig. 2d). Hence, we believe that there must be an optimum surface density of LaB$_6$ to maximize the yield, as well as the rate of production of the as-synthesized CNTs.

The formation LaB$_6$ decorated CNTs (as discussed above) from the bulk graphite is an extremely slow and dynamic sloid state transformation process, which can probably be understood in detail provided all the intermediate stages of the process are duly characterized. However, finding the actual physicochemical mechanism behind our observations is time consuming and beyond the scope of present communication.

4. **Conclusions**

In conclusion, for the first time we have discovered a quite simple, economical, pollution free and highly reproducible way to convert commercially available polycrystalline bulk graphite into LaB$_6$ decorated MWCNTs spontaneously at normal atmospheric conditions with high conversion ratio. The synthesis is a batch process whose yield is proportional to the surface area of the feedstock graphite. To activate the conversion, the feedstock graphite with some commercially available LaB$_6$ microcrystalline particles must be eroded partially by an electric arc-plasma preferably in pure nitrogen atmosphere and then left to the atmosphere. Other rare-earth hexaborides and mixed rare-earth hexaborides may also be used to activate the process. The conversion typically starts after a couple of hours of arcing and may continue for months without requiring any additional energy feed. Hence, the process shows promises for developing a new way to harvest LaB$_6$ decorated CNTs at a commercial level. Most of the CNTs generated by this process are several microns long with typically 10 – 15 nm diameter and sharp conical caps. Owing to the typical morphology and structure of these CNTs, they can act as efficient field emitters.

**Conflicts of interest**

The authors declare no conflicts of interest.

**Acknowledgements**

Shalaka Kamble acknowledges the Department of Science and Technology (DST), Government of India for providing financial support through Women Scientist Scheme-A (SR/WOS-A/PM-110/2018). V. L. Mathe acknowledges BRNS, Mumbai, India for the project grant (Grant No BRNS-34/14/23/2015/34019).

**Declaration of Generative AI and AI-assisted technologies in the writing process**

The authors declare that no AI and AI-assisted technologies have been used to draft the manuscript.